\documentclass[traditabstract]{aa}
\usepackage{graphicx}
\usepackage{txfonts}
\begin{document}
   \title{The counterpart/s of IGR J20159+3713/SWIFT J2015.9+3715: dissecting a complex region with 
emission from keV to TeV}

%   \subtitle{The counterpart/s of IGR J20159+3713/SWIFT J2015.9+3715}

    \titlerunning{The counterpart/s of IGR J20159+3713/SWIFT J2015.9+3715}
\authorrunning{L.~Bassani}

   \author{L. Bassani\inst{1}, R.~Landi\inst{1}, A. Malizia\inst{1}, J. B. Stephen\inst{1},
A. Bazzano\inst{3},
A. J. Bird\inst{4}, and P. Ubertini\inst{3}}
   \offprints{bassani@iasfbo.inaf.it}
\institute{INAF/IASF Bologna, via Piero Gobetti 101, I--40129 Bologna, Italy 
\and
INAF/IAPS Rome, Via Fosso del Cavaliere 100, I--00133 Rome, Italy \and
School of Physics and Astronomy, University of Southampton, Highfield, SO17 1BJ, UK}

   \date{Received  / accepted}
% \abstract{}{}{}{}{} 
% 5 {} token are mandatory

\abstract {
We report on the identification of a new soft gamma-ray source, namely IGR J20159+3713/SWIFT 
J2015.9+3715, first detected by \emph{INTEGRAL}/IBIS and then confirmed by \emph{Swift}/BAT. The source, 
which has an observed 20--100 keV flux in the range $(0.7-1.4) \times 10^{-11}$ erg cm$^{-2}$ 
s$^{-1}$, encloses a 
\emph{Fermi} variable source (2FGL J2015.6+3709) and is spatially close to a TeV emitter (VER J2016+372). 
Thanks to X-ray follow-up observations performed with the X-ray telescope on board \emph{Swift}, we have 
been able to identify the new IBIS/BAT detection with the combined emission of the blazar B2013+370 and 
the cataclysmic variable RX J2015.6+3711. Both objects show variability in X-rays, with the CV being the 
most variable of the two. At high energies (above 20 keV) the emission is likely dominated by B2013+370, 
but the contribution from RX J2015.6+3711 is not negligible. The blazar emits up to GeV frequencies 
where 
it is seen by \emph{Fermi}, while the cataclysmic variable has a bremsstrahlung temperature which is too 
low to provide any contribution at these high energies. These findings also indicate that the 
\emph{INTEGRAL}/\emph{Swift} source is not associated with the TeV emission, which is most likely due to 
the supernova remnant (SNR)/pulsar wind nebula (PWN) CTB 87.
}


\maketitle 

\section{Introduction}

IGR J20159+3713/SWIFT J2015.9+3715 is a new soft gamma-ray source first detected with the IBIS imager on 
board the \emph{INTEGRAL} satellite (Krivonos et al. 2010) and then confirmed by the BAT instrument on 
board the \emph{Swift} mission (Cusumano et al. 2010). The source lies in a region of complex gamma-ray 
emission. In fact, IGR J20159+3713/SWIFT J2015.9+3715 is close to a GeV/\emph{Fermi}/EGRET source (3EG 
J2016+3657/1FGL J2015.7+3708/2FGL J2015.6+3709) associated with the blazar B2013+370 (Halpern et al. 
2001); the GeV emission is variable supporting the hypothesis of an extragalactic origin for the 
\emph{Fermi} detection (Kara et al. 2012). Close by there is also a newly discovered TeV/VERITAS object 
(VER J2016+372), whose position is compatible with that of CTB 87 a SNR/PWN 
located at 6 kpc in the Perseus arm, but in the line of sight of Cyg OB1 (Aliu et al. 2011). The 
properties and energetics of the underlying pulsar are poorly known although, recently, some information 
have been obtained by means of \emph{Chandra} observations (Matheson, Safi-Harb \& Kothes 2013). The 
association between the VERITAS and \emph{Fermi} source seems to be excluded at the 99\% confidence level 
(c. l.) as reported by Aliu et al. (2011), while the connection of these GeV and TeV sources with the 
IBIS/BAT detection has never been discussed before in the literature. To complicate the situation, we also 
note that IGR J20159+3713/SWIFT J2015.9+3715 has been identified with different source typology in the 
various hard X-ray catalogues in which it is reported: as a SNR by Krivonos et al. (2010; 2012), as a 
cataclysmic variable (CV) in Cusumano et al. (2010), and as a blazar in the latest BAT 70-month survey 
(Baumgartner et al. 2013). This clearly reflects the complexity of this sky region and points to a 
reanalysis of all the available observations to understand the nature of the IBIS/BAT source and its 
relation to the nearby GeV/TeV emitters.

Through a full analysis of these observations, we will demonstrate that the IBIS/BAT emission is 
likely due to the contribution of both the CV and the blazar and is unrelated to the SNR/PWN.

\section{Gamma-ray emission in and around IGR J20159+3713/SWIFT J2015.9+3715}

\begin{figure} 
\centering
\includegraphics[width=1.\linewidth,angle=0]{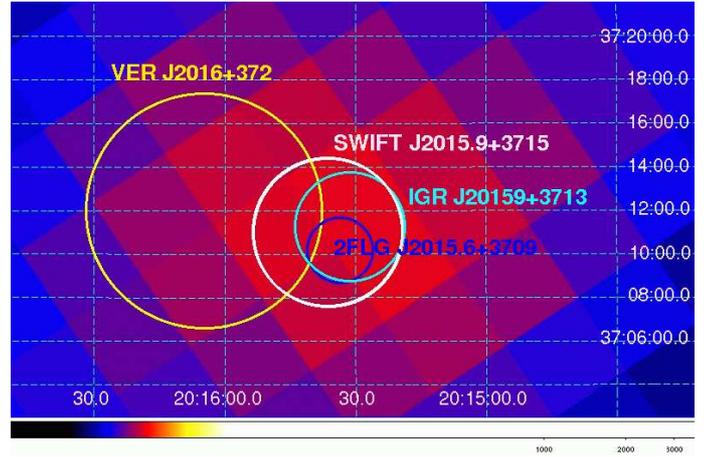}
\caption{IBIS 17--60 keV significance map containing IGR J20159+3713/SWIFT J2015.9+3715. The white 
circle corresponds to the IBIS positional uncertainty, while the light blue is that from BAT 
70-month survey; the small dark blue circle shows the positional uncertainty of 
the \emph{Fermi} source, while the large yellow circle represents the area where the VERITAS emission 
comes from.}
\label{fig1}
\end{figure}

Figure~\ref{fig1} is a cut-off image of the region surrounding IGR J20159+3713/SWIFT J2015.9+3715 as 
seen by \emph{INTEGRAL}/IBIS in the 17--60 keV band\footnote{available at:\\ 
http://hea.iki.rssi.ru/integral/nine-years-galactic-survey/index.php.}. It shows the 9-year average 
IBIS significance map with superimposed the source positional uncertainty (Krivonos et al. 
2012); for comparison we also show the error circle reported in the latest BAT 70-month catalogue 
(Baumgartner et al. 2013). The source is detected by IBIS with a significance of $\sim$12$\sigma$ at 
a position corresponding to R.A.(J2000) = $20^{\rm h}15^{\rm m}31^{\rm s}.44$ and Dec.(J2000) = 
$+37^{\circ}11^{\prime}16^{\prime\prime}.8$, with an associated uncertainty of $\sim 3.^{\prime}4$ 
(90\% c. l.). The BAT detection is at 8$\sigma$ c. l. and its positional uncertainty is slightly 
larger but fully compatible with the IBIS one. The location of IGR J20159+3713/SWIFT J2015.9+3715 is 
1.20 degrees above the Galactic plane, making difficult to discriminate whether it is a Galactic or 
an extragalactic object. A simple power law provides a good fit to the high-energy data and a photon 
index $\Gamma\sim2$ combined to an observed 20--100 keV flux in the range ($0.7-1.4) \times 
10^{-11}$ erg cm$^{-2}$ s$^{-1}$ (Krivonos et al. 2012, Cusumano et al. 2010, Baumgartner et al. 
2013).

Here and in the following, spectral analysis is performed with XSPEC v.12.8.0 package and errors are 
quoted at 90\% c. l. for one interesting parameter ($\Delta\chi^{2}=2.71$). 

As mentioned in the introduction, this hard X-ray source has different optical counterparts in the 
different catalogues in which it is listed, suggesting that either some associations are wrong or that 
the emission is from multiple objects.
  
The GeV/TeV detections reported recently in the literature are also plotted for comparison in 
Figure~\ref{fig1}. For \emph{Fermi} we used the position and relative uncertainty quoted by Kara et al. 
(2012) and based on 31 months of \emph{Fermi} data; for the VERITAS detection we used the position 
reported by Aliu et al. (2011), taking as error the integration radius (0.09 degrees) employed to search 
for the TeV signal. As evident in Figure~\ref{fig1}, spatial considerations indicate a possible 
association between emissions in different wavebands, especially between hard X-ray and GeV/gamma-ray 
energies; it also emphasises the inability of current high-energy instruments to resolve a complex 
region such as this one.

The time-averaged \emph{Fermi} spectrum is well described by a power law with photon index $\Gamma = 
(2.57\pm0.02)$ and a flux above 1 GeV of $9.3 \times 10^{-9}$ cm$^{-2}$ s$^{-1}$; the 
\emph{Fermi} spectrum is slightly softer than the one found previously by EGRET 
\big($\Gamma = (2.09\pm0.11)$, Hartman et al. 1999\big). Analysis of the \emph{Fermi} light curve 
indicates flux variability and the 
presence of two flares in 2009 and 2010 (Kara et al. 2012). The presence of variability in the GeV 
emission, a property which is typical of blazars, clearly points to an association of the \emph{Fermi} 
source with B2013+370, a flat spectrum radio source which is still optically unclassified. On the basis 
of the source overall properties, Kara et al. (2012) concluded that B2013+370 is most likely a low peaked 
BL Lac or a flat spectrum radio quasar (FSRQ).

Above 1 TeV, the emission from VER J2016+372 is $\sim$0.8--1\% that of the Crab Nebula and the 
spectrum is best fitted with a power law having a photon index $\Gamma = (2.1\pm0.5)$; the 
measured spectrum and the absence of variability observed throughout the TeV observation are 
properties similar to those of PWNs previously detected at TeV energies (Aliu et al. 2011). For this 
reason, the source has been associated with CTB 87, a SNR/PWN complex located at the centre of the TeV 
error circle, while its relation to B2013+370 has been considered unlikely.

\begin{figure} 
\centering
\includegraphics[width=1.\linewidth,angle=0]{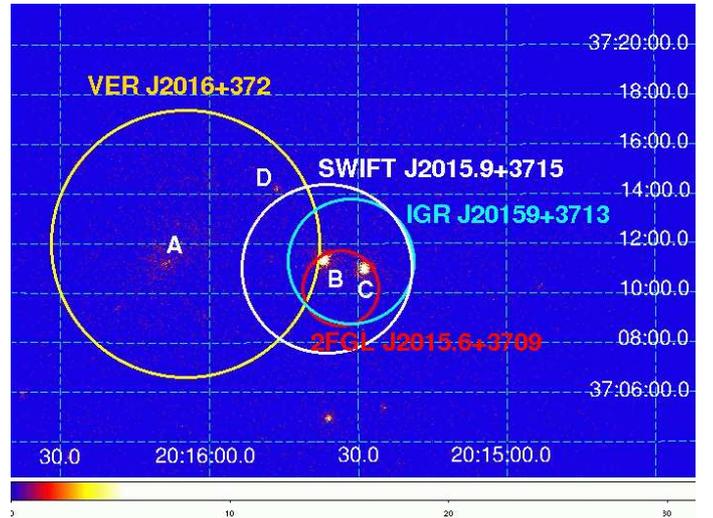}
\caption{XRT 0.3--10 keV count map of the region surrounding IGR J20159+3713/SWIFT 
J2015.9+3715; error circles are the same as in Figure~\ref{fig1}.}
\label{fig2}
\end{figure}

\section{\emph{Swift}/XRT follow-up observations}

\subsection{Imaging analysis}

In order to disentangle the emission from various sources and in different wavebands, a useful tool 
is to analyse the region at softer X-ray energies, where the instrument angular resolution and point 
source location accuracy are generally far better than at higher energies. To this purpose we used the 
\emph{Swift}/XRT instrument, which has an angular resolution of 18 arcseconds and a source location 
accuracy of a few arcsec (Burrows et al. 2005). Specifically, we used observations performed during the 
periods November 2006 and August 2010: eight pointings of reasonable exposure were carried out during 
these periods for a total time of $\sim$30 ks. XRT data reduction was performed using the XRTDAS standard 
data pipeline package ({\sc xrtpipeline} v. 0.12.6), in order to produce screened event files. All data 
were extracted only in the Photon Counting (PC) mode (Hill et al. 2004), adopting the standard grade 
filtering (0--12 for PC) according to the XRT nomenclature.

\begin{table*}
\begin{center}
\centering
\footnotesize
\caption{\emph{Swift}/XRT detections.}
\label{tab1}
\begin{tabular}{lcccccccc}
\hline
\hline
Source  &    \multicolumn{3}{c}{Position}       &  Extension    &   \multicolumn{2}{c}{$\sigma$}      &   Association    &  Type   \\
              &    R.A.(J2000)  &  Dec.(J2000)  & error &              &     &    & \\
    &    &   &   (arcsec)   &   &  (0.3--10 keV) & ($>$ 4 keV) &   &       \\
\hline
\hline
A$^{a}$       & $20^{\rm h}16^{\rm m}08^{\rm s}.70$    &  $+37^{\circ}11^{\prime} 14^{\prime\prime}.10$  
& 6.$^{\prime\prime}$00 & extended   & 17 &  7.9  &  CTB 87         
& SNR/PWN \\
B             & $20^{\rm h}15^{\rm m}36^{\rm s}.95$   &  $+37^{\circ}11^{\prime} 21^{\prime\prime}.47$ & 3.$^{\prime\prime}$56 & point-like & 37 &  
22   & RX J2015.6+3711 
&  CV     \\
C             & $20^{\rm h}15^{\rm m}28^{\rm s}.76$   &  $+37^{\circ}10^{\prime} 58^{\prime\prime}.87$ & 3.$^{\prime\prime}$57 & point-like & 36 &  19.5 & B2013+370      
& Blazar  \\
D             & $20^{\rm h}15^{\rm m}46^{\rm s}.20$    &  $+37^{\circ}14^{\prime} 13^{\prime\prime}.20$  
& 6.$^{\prime\prime}$00 & point-like & 8  & 2.5   & 
X201546.42+371413.4 & unid \\
\hline
\hline
\end{tabular}
\begin{list}{}{}
\item $^{a}$ The position reported here coincides, within uncertainties, with that of the bright spot detected 
by \emph{Chandra} within the SNR/PWN diffuse emission and is
associated with the X-ray source CXOU J201609.2+371110, firstly detected by Matheson, Safi-Harb \& Kothes 
(2013).
\end{list}
\end{center} 
\end{table*}

Data from all eight observations were combined together in a single 0.3--10 keV count map which 
is displayed in Figure~\ref{fig2}, together with all relevant error circles. The map clearly shows 
the presence of four X-ray sources, three brighter ones (A, B, and C) and one dimmer (D). 
Table~\ref{tab1} lists for each of these detections, the XRT position and uncertainty, the source 
extension, the detection significance in the 0.3--10 keV energy band and above 4 keV, the source 
association and its nature.

The first source (A) is extended and the position of its brightest spot is compatible with the peak of 
the X-ray emission associated with the PWN recently discovered in CTB 87 (Safi-Harb 2012; Matheson, 
Safi-Harb \& Kothes 2013). The source morphology is similar, although less detailed, to what is evident 
in a \emph{Chandra} image of the region where both a compact nebula (with a torus/jet-like structure) and 
a diffuse nebula (with a cometary-like morphology) are clearly visible (Safi-Harb 2012; Matheson, 
Safi-Harb \& Kothes 2013 ).

The second object (B) is associated with RX J2015.6+3711, one of the 14 bright ROSAT sources found 
within 3EG J2016+3657 error box and discussed in detail by Halpern et al. (2001). In their paper, RX 
J2015.6+3711 is source N.2 and its optical spectrum (display for the first time in Figure 5 of Hapern et 
al. 2001) is that of a cataclysmic variable, probably of the magnetic type since its He II j4686 emission 
line is as strong as Hb j4861; the spectrum is also reddened, implying a source distance greater than 1 
kpc.

The third source (C) is coincident with the blazar B2013+370. Note that the distance between the CV and 
the blazar is only 1.6 arcmin, thus making the two unresolved at high energies.

Finally, the last much dimmer object coincides with a \emph{Chandra} source, but has no clear 
association at other wavelengths (Ptak \& Griffiths 2003).

A few notes can be made on the basis of Figure~\ref{fig2}: IBIS error circle excludes the SNR/PWN CTB 
87, thus implying a wrong association in the Krivonos et al. (2010; 2012) catalogues. \emph{Fermi} error
circle is interestingly compatible with both BAT 70-month and IBIS error circles, suggesting some 
association and highlighting the possibility that beside the blazar also the CV could be considered a 
less likely but still a possible GeV emitter. Finally, within the VERITAS positional uncertainty the 
SNR/PWN remains the most likely candidate, although source D cannot be totally disregarded.

Imaging this region of the sky above 4 keV could be of some help: as shown in Table~\ref{tab1}, the CV 
and the blazar are still the brightest objects detected at these higher energies (at 22 and 19.5$\sigma$, 
respectively), 
the SNR/PWN is still seen but only at 7.9$\sigma$, while the fourth object dissapears below the 
2.5$\sigma$ detection level. While this evidence does not allow a clear identification of the \emph{Fermi} 
and VERITAS counterparts, it certainly helps to clarify which is the real association to the IGR 
J20159+3713/SWIFT J2015.9+3715 source. Simply on the basis of spatial and brightness considerations, both 
the CV and the QSO are the most likely counterparts of the hard X-ray detection. Further clues may 
come from spectroscopic considerations as well as variability studies.

\subsection{Spectral analysis}

Events for spectral analysis were extracted within a circular region of radius 20$^{\prime \prime}$ 
(which encloses about 90\% of the PSF at 1.5 keV, Moretti et al. 2004) centred on the position of each 
interesting source. The background was extracted from various source-free regions close to the X-ray 
source using both circular/annular regions with different radii, in order to ensure an evenly sampled 
background. In all cases, the spectra were extracted from the corresponding event files using {\sc 
XSELECT} software and binned using {\sc grppha} in an appropriate way, so that the $\chi^{2}$ statistic 
could be applied. We used the latest version (v.013) of the response matrices and create individual 
ancillary response files (ARF) using {\sc xrtmkarf v.0.6.0}. In all our fitting procedures we have used a 
Galactic column density which in the direction of these sources is $1.15\times 10^{22}$ cm$^{-2}$ 
(Kalberla et al. 2005).

Table~\ref{tab2} reports for each X-ray measurement, the observation date, the relative exposure, the 
net count rate in the 0.3--10 keV energy band and the detection significance for the CV RX 
J2015.6+3711, the blazar B2013+370 and the SNR/PWN CTB 87. There is clear evidence that all sources 
underwent some flux changes over the XRT monitoring: the reduced $\chi^{2}$ (for a constant model at 
the mean) for sources A, B and C is 15, 50, 20, respectively, which indicates that all 
three objects changed flux in time with the CV being the most variable of the three. Variability in 
blazars, CV and even in PWN is not unexpected and can be used to characterise further these objects.

\begin{table*}[t]
%\begin{center}
\centering
\footnotesize
\caption{Log of the \emph{Swift}/XRT observations and count rates of each detection.}
\label{tab2}
\begin{tabular}{lcccccc}
\hline
\hline
 Observation    &     Date      &  Exposure$^{a}$   &   \multicolumn{3}{c}{Count Rate$^{b}$}   \\
                &               &                   &   CV    &   Blazar   &  PWN$^{c}$   \\
                &               &     (sec)         &   (10$^{-3}$ counts s$^{-1}$)       &    (10$^{-3}$ counts s$^{-1}$)  & (10$^{-3}$ counts s$^{-1}$) \\
\hline
\hline
\#1             &  July 12 2006   &   1528          &  40.1$\pm$ 5.2      & 11.3$\pm$ 2.9    & $1.3\pm0.9$   \\
\#2             &  Nov  12 2006   &   4490          &  46.7$\pm$ 3.3      & 22.6$\pm$ 2.3    & $1.4\pm0.7$   \\
\#3             &  Nov  17 2006   &   7367          &  66.8$\pm$ 3.0      & 24.7$\pm$ 1.9    & $13.9\pm1.4$  \\
\#4             &  Nov  24 2006   &    170          &  87.6$\pm$ 24.5     & 27.8$\pm$ 13.3   & $10.6\pm8.4$   \\
\#5             &  Aug   5 2010   &    962          &  57.9$\pm$ 8.0      & 69.3$\pm$ 8.7    & $7.9\pm3.3$    \\
\#5             &  Aug   6 2010   &   7307          &  43.5$\pm$ 2.5      & 55.6$\pm$ 2.8    & $9.1\pm1.2$    \\
\#6             &  Aug  22 2010   &   4310          &  43.3$\pm$ 3.2      & 73.3$\pm$ 4.2    & $8.8\pm1.5$    \\
\#8             &  Aug  30 2010   &   4146          &  32.4$\pm$ 2.9      & 46.4$\pm$ 3.4    & $6.8\pm1.4$    \\             
\hline
                &     Total       &  30280          &  48.4$\pm$ 1.3      & 42.6$\pm$ 1.2    & $10.2\pm0.6$   \\
\hline
\hline
\end{tabular}
\begin{list}{}{}
\item $^{a}$ Total on-source exposure time;
\item $^{b}$ The count rate is estimated in the 0.3--10 keV energy range;
\item $^{c}$ The PWN count rate has been extracted from a circle of 20$^{\prime\prime}$
radius centred on the brightest X-ray spot.
\end{list}
%\end{center} 
\end{table*}

Unfortunately, it also complicates the picture, expecially in the case of B2013+370 and RX J2015.6+3711,
where it becomes quite difficult to disintangle the contribution of one source from the other at high 
energies purely on the basis of flux variations. Spectral information can be useful, in this respect and 
therefore, to enhance the signal to noise ratio and obtain good constrain on the source parameters, we 
have combined XRT data from two-epoch observations (2006 and 2010) to obtain two states of the blazar 
flux, while all CV data have been added together to study the source average X-ray spectrum.

\begin{figure} 
\centering
\includegraphics[width=1.\linewidth,angle=0]{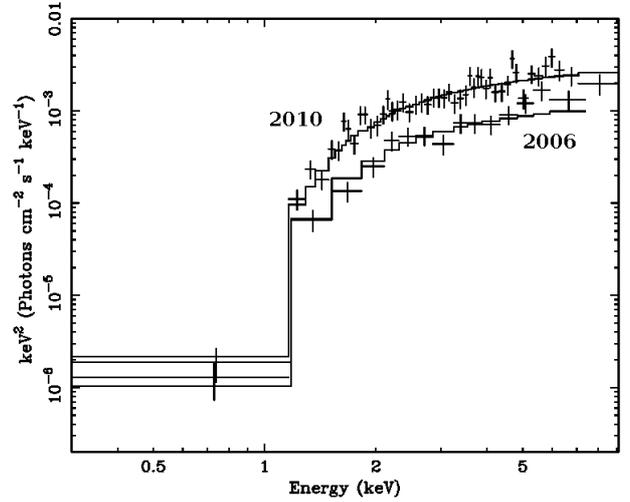}
\caption{Two-epoch XRT spectra of the blazar B2013+370 fitted simultaneously with an absorbed power law.}
\label{fig3}
\end{figure}

The two blazar spectra have been fitted with the same model, i.e. an absorbed power law with different 
1 keV normalisations; this is justified by the fact that within uncertainties Kara et al. (2012) report 
a similar XRT spectum for the 2006 and 2010 periods. The absorption is partially Galactic and partially 
intrinsic to the source. The best fit \big($\chi^{2}/\nu = 66.4/59$\big) is obtained with a photon 
index of $\Gamma=\big(1.66^{+0.22}_{-0.20}\big)$ and an intrisic column density of 
\big($0.78^{+0.33}_{-0.29}\big)\times 10^{22}$ cm$^{-2}$; the 2--10 keV flux is $\sim 2 \times 
10^{-12}$ erg cm$^{-2}$ s$^{-1}$ and $\sim 5 \times 10^{-12}$ erg cm$^{-2}$ s$^{-1}$ for the 2006 and 
2010 period, respectively (see Figure~\ref{fig3})\footnote{Note that in Kara et al. (2012) the 2006 
flux is brighter than the 2010 one, possibly due to a misprint.}. The extrapolation of these two spectra 
to high energies provides a 20--100 keV flux in the range $(0.5-1.2) \times 10^{-11}$ erg cm$^{-2}$ 
s$^{-1}$.

Note that the addition of an intrinsic column density provides a significant improvement in the fit ($> 
99.99\%$ c. l. using the F-test). Intrinsic absorption is a common property of 
radio-loud/high-redshift blazars (Page et al. 2005) and it is generally ascribed to the presence of 
absorbing material 
in the jet. An alternative possibility is that the continuum is intrinsically curved (Tavecchio et al. 
2007) as expected if the emission is described by a broken power law: however, this model does not fit 
the data properly leaving absorption as the only viable scenario. Overall, we can conclude that the XRT 
spectra of B2013+370 indicate different source fluxes over time and also suggest that this object can 
provide a large fraction of the flux seen above 20 keV.

\begin{figure} 
\centering
\includegraphics[width=1.\linewidth,angle=0]{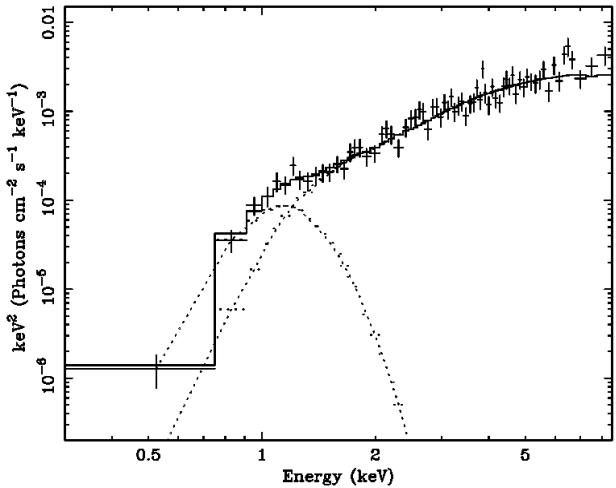}
\caption{Best-fit spectrum of the CV RX J2015.6+3711 (see details in the text).}
\label{fig4}
\end{figure}

The soft X-ray spectrum of the CV is more difficult to characterise. As reported by Landi et al. 
(2009), magnetic CVs are generally described by various spectral components, more specifically a black 
body model plus an absorbed bremsstrahlung one, both passing through the Galactic column density. The 
absorption related to the bremsstrahlung component covers the source only partially (pcfabs model in 
XSPEC). Such model, if applied to the average XRT spectrum of RX J2015.6+3711, is feasible, but 
provides a bremsstrahlung temperature which is too high ($kT\ge 70$ keV). Ideally, one needs to have 
broad-band coverage for such a detailed analysis, but this is difficult for RX J2015.6+3711 given the 
significant contamination from the blazar at high energies. One solution is to assume that the BAT/IBIS 
emission is due to the combined contribution of two objects, i.e. the blazar, characterised by a simple 
power law and the CV, described by a bremsstrahlung component. By doing this and letting the photon index 
of the blazar power law vary within the observed boundaries (1.44--1.88), it is possible to put more 
reasonable constraints on the CV bremsstrahlung temperature, which falls in the range 13--15 
keV, i.e. closer to the values typically observed in this type of binary systems (Landi et al. 2009). 
This information can in turn be used to describe the CV average spectrum seen by XRT. Adopting the 
complex model described above (i.e. black body plus partially absorbed bremsstrahlung both passing 
through the Galactic column density), a good fit \big($\chi^{2}/\nu = 70.5/65$\big) is obtained with a 
black body temperature of \big(0.121$^{+0.29}_{-0.20}$~\big) keV and a column density of 
\big($8.8^{+6.1}_{-4.1}\big)\times 10^{22}$ cm$^{-2}$ covering \big($0.64^{+0.07}_{-0.06}\big)\%$ of 
the source (see Figure~\ref{fig4}); the 2--10 keV flux is $\sim 4.4 \times 10^{-12}$ erg cm$^{-2}$ 
s$^{-1}$, while its extrapolation to high energies gives a 20--100 keV flux of $\sim$$(2.1-2.6) \times 
10^{-12}$ erg cm$^{-2}$ s$^{-1}$. A closer inspection of Figure~\ref{fig4} indicates the presence of  
excess emission around 6.4 keV. The addition of an iron line is required by the data at 99$\%$ c. l. and 
provides an energy centroid $E = \big(6.48^{+0.10}_{-0.08}$\big) keV and an equivalent width $EW = 
\big(480^{+210}_{-246}$~\big) eV.

The overall picture which therefore emerges from the analysis of the XRT spectral data leads one to 
conclude that IGR J20159+3713/SWIFT J2015.9+3715 is likely the unresolved emission of two nearby 
high-energy emitters, the hard X-ray blazar B2013+370 and the likely magnetic CV RX J2015.6+3711. 
Purely on the basis of spectral information and data extrapolation, we estimate that the former 
contributes around 80--90$\%$ of the hard X-ray flux, while the latter provides the remaining 
10--20$\%$ of 20--100 keV photons.

For the sake of completeness, we have also performed the spectral analysis of the X-ray emission 
associated with CTB 87, which has recently been studied in details by Matheson, Safi-Harb \& Kothes 
(2013) exploiting \emph{Chandra} observations. Also in this case all XRT data have been added 
together to study the average X-ray spectrum of this source. Unfortunately, due to the lower XRT 
spatial resolution, we cannot perform a space-resolved analysis with the same accuracy of 
\emph{Chandra}, but have nevertheless been able to study the spectral properties of the entire system 
as well as of the central compact region. The spectrum of the entire object is well fitted with a 
simple power law absorbed by Galactic absorption and is characterised by the following parameters: a 
photon index $\Gamma = (1.78\pm0.15$) and an observed 2--10 keV flux of $\sim$ $2\times10^{-12}$ erg 
cm$^{-2}$ s$^{-1}$. This is fully compatible, within uncertainties, with the results found by 
Matheson, Safi-Harb \& Kothes (2013) using \emph{Chandra} data (see total diffuse emission in their 
Table 2). Restricting the analysis to a region of 20 arcsec around the brightest XRT pixel, we obtain 
instead a slightly flatter spectrum ($\Gamma = 1.42\pm0.22$) and a lower 2--10 keV flux ($\sim 8 
\times 10^{-13}$ erg cm$^{-2}$ s$^{-1}$). Our central region includes the putative pulsar, the 
compact nebula, and the inner diffuse emission region resolved by \emph{Chandra} (Matheson, Safi-Harb 
\& Kothes 2013); also in this case our spectral parameters compare well with those of \emph{Chandra} 
despite the lack of the same resolution in our data. In particular, we confirm the flattening of the 
X-ray spectrum when only the central part of CTB 87 is considered. As a final remark, we note that 
Matheson, Safi-Harb \& Kothes (2013) used two \emph{Chandra} observations which they merged together 
to enhance the signal to noise ratio, thus missing information on variability. Given enough 
statistics, it would be interesting to explore which of the various features detected by 
\emph{Chandra} in the source inner region is responsible for the flux variations seen by XRT.

\section{Connection to the GEV/TeV emission}

Starting from the above discussion, it is possible to move forward and assess the relation between 
IGR J20159+3713/SWIFT J2015.9+3715 and the GeV/TeV emitters. The IBIS/BAT emission, being mostly due 
to the blazar, is clearly connected to the GeV object. The SED of B2013+370 is shown and discuss by 
Kara et al. (2012): data above 20 keV are potentially interesting as they can provide a way to 
discriminate between different emission scenarios. Using the blazar contribution estimated in the 
present paper, we favour a scenario where the source is powered by a leptonic model with the 
synchrotron self Compton (SSC) component dominating at X-rays and the external Compton (EC) one 
providing the gamma output. More specifically, the data above 20 keV are significantly below the 
SSC+EC2 model favoured by Kara et al. (2012) (i.e. that with a lower temperature external field), and 
are instead more compatible with the SSC+EC1 model (i.e. that with a higher temperature field). We 
also note that the IBIS/BAT spectrum is relatively flat and more compatible with that expected in the 
case of the SSS+EC1 model. Clearly, variability and source contamination is an issue in this case and 
for a more precise analysis of the source SED not only simultaneous data are necessary, but also 
better angular resolution at the highest X-ray energies. Nevertheless, it is interesting to underline 
here the importance of hard X--ray information for such type of studies.

As for the CV contribution to the \emph{Fermi} source, this is very likely negligible, if not null, 
although one must remember the case of XSS J12270--4859. This object has also been detected at hard 
X-ray energies by \emph{INTEGRAL}/IBIS and \emph{Swift}/BAT, it has then be classified as a CV of the 
magnetic type (Masetti et al. 2006) and subsequently it was associated with a \emph{Fermi} source 
(1FGL J1227.9--4852/2FGL J1227.7--4853) listed in the first and second catalogues (Abdo et al. 
2010;2012); only recently, it was described as a peculiar type of X-ray binary (De Martino et al. 
2010, De Martino et al. 2013; Hill et al. 2011). Furthermore, XSS J12270--4859 is not the first time 
that a CV has been associated with a GeV emitter. PSR J1023+0038 is another GeV emitter initially 
believed to be the first radio selected CV (Thorstensen \& Armstrong 2005; Bond et al. 2002) and only 
after closer scrutiny classified as a pulsar, in much the same way as XSS J12270--4859 was initially 
described as a CV and then found to be a peculiar object. This is simply to caution the reader that 
although RX J2015.6+3711 is an unlikely contributor to the \emph{Fermi} source, it should be studied 
more in depth before it can be totally disregarded as a GeV emitter. This is even more true in view 
of its X-ray flux variability, which is clearly present in the XRT data. The scenario at TeV energies 
is by far simplier as the VERITAS TeV source is most likely associated with CTB 87 and related either 
to the SNR, its associated pulsar/PWN or both. In this case, no association is implied with the 
\emph{INTEGRAL}/\emph{Swift} high energy object.

\section{Conclusions}

In this paper we have discussed the nature of the newly discovered \emph{INTEGRAL}/\emph{Swift} source 
IGR J20159+3713/SWIFT J2015.9+3715 which lies in a complex region of GeV/TeV emission.

Through X-ray follow-up observations with \emph{Swift}/XRT, we have been able to identify this new
detection with the combined emission of the blazar B2013+370 and the CV RX J2015.6+3711. Both objects are 
variable in X-rays, with the CV being the most variable of the two. The emission 
above 20 keV is likely dominated by B2013+370, but the contribution from RX J2015.6+3711 is not small. 
The blazar emits up to GeV frequencies where it is seen by \emph{Fermi}, while the CV temperature is too 
low to provide any contribution at these high energies. Finally, the \emph{INTEGRAL}/\emph{Swift} source 
is not associated with the TeV emission which is most likely due to the SNR/PWN CTB 87.

\begin{acknowledgements} 

We acknowledge finacial support from ASI under contract ASI I/033/10/0 and ASI/INAF I/009/10/0. 
This research has made use of  the  NASA/IPAC Extragalactic Database (NED) operated by 
Jet Propulsion Laboratory (California Institute of Technology) and of the HEASARC 
archive provided by NASA's Goddard Space Flight Center. The authors also acknowledge the use of public 
data from the Swift data archive.

\end{acknowledgements}


\begin{thebibliography}{}

\bibitem{} Abdo, A. A., Ackermann, M., Ajello, M. et al. 2012, ApJS, in press, arXiv:1108.1435
\bibitem{} Abdo, A. A., Ackermann, M., Ajello, M., et al. 2010, ApJS, 188, 405
\bibitem{} Aliu,E.,  et al. E. 2011, Proceedings of the 32nd ICRC, Beijing, China, arXiv:1110.4656
\bibitem{} Baumgartner, W. H., Tueller, J., Markwardt, C. B., et al. 2013, submitted to ApJS,
arXiv:1212.3336 
\bibitem{} Bond, H. E., White, R. L., Becker, R. H., O'Brien, M. S. 2002, PASP, 114, 1359
\bibitem{} Burrows D. N., Hill, J. E.,  Nousek, J. A.,  et al. 2005, Space Sci. Rev., 120, 165
\bibitem{} Cusumano, G., La Parola, V., Segreto, A., et al. 2010, A\&A, 510, A48
\bibitem{} De Martino, D., Falanga, M., Bonnet--Bidaud, J.--M., et al. 2010, A\&A, 515, A25
\bibitem{} De Martino, D., Belloni, T., Falanga, M., et al. 2013, A\&A, 550, A89
\bibitem{} Halpern, J. P., Eracleous, M., Mukherjee, R., Gotthelf, E. V. 2001, ApJ, 551, 1016
\bibitem{} Hartman, R. C., Bertsch, D. L., Bloom, S. D., et al. 1999, ApJS, 123,79
\bibitem{} Hill, A. B., Szostek, A., Corbel, S., et al. 2011, MNRAS 415, 235
\bibitem{} Hill J. E., Burrows, D. N., Nousek, J. A., et al. 2004, Proc. SPIE, 5165, 217
\bibitem{} Kalberla P. M. W., Burton W. B., Hartmann D., et al., 2005, A\&A, 440, 775
\bibitem{} Kara, E., Errando, M., Max-Moerbeck, W., et al. 2012, ApJ, 746, 159
\bibitem{} Krivonos, R., Tsygankov, S., Lutovinov, A., et al., 2012, A\&A, 545, A27
\bibitem{} Krivonos, R., Tsygankov, S., Revnivtsev, M., et al. 2010, A\&A, 523, A61
\bibitem{} Landi, R., Bassani, L., Dean, A. J., et al. 2009, MNRAS, 392, 630 

\bibitem{} Masetti, N., Morelli, L., Palazzi, E., et al. 2006, A\&A, 459, 21
\bibitem{} Matheson, H., Safi-Harb, S. \& Kothes, R. 2013, ApJ, in press, arXiv:1307.1084  
\bibitem{} Moretti, A., Campana, S., Tagliaferri, G., et al. 2004, Proc. SPIE, 5165, 232
\bibitem{} Page, K. L., Reeves, J. N., O'Brien, P. T., Turner, M. J. L. 2005, MNRAS, 364, 195
\bibitem{} Ptak, A. \& Griffiths, R. 2003, Astronomical Data Analysis Software and Systems XII ASP 
Conference Series, Vol. 295, H. E. Payne, R. I. Jedrzejewski, and R. N. Hook, eds., p. 465
\bibitem{} Safi-Harb, S. 2012, Proc. of the 5th International Meeting on High Energy Gamma-Ray Astronomy. 
AIP Conference Proceedings, Vol. 1505, p. 13
\bibitem{} Tavecchio, F., Maraschi, L., Ghisellini, G., Kataoka, J. 2007, ApJ, 665, 980
\bibitem{} Thorstensen, J. R. \& Armstrong, E. 2005, 2005, AJ, 130, 759
\end{thebibliography}
\end{document}